\documentclass[english,10pt]{article}
\usepackage[english]{babel}
\usepackage[applemac]{inputenc}
\usepackage{amsmath}
\usepackage{amsfonts}
\usepackage{amssymb}
\usepackage{graphicx}
\usepackage{tabularx}
\usepackage{geometry}
\usepackage{harvard}
\usepackage{multicol}
\newcommand{\E}{\mathbb{E}}
\newcommand{\V}{\textrm{Var}}
\newcommand{\R}{\mathbb{R}}

\newtheorem{theoreme}{Theorem}

\makeatletter

\newenvironment{figurehere}
  {\def\@captype{figure}}
  {}
\makeatother

\makeatletter

\makeatletter
\makeatletter

\makeatletter

\makeatletter
\renewcommand\section{\@startsection {section}{3}{\z@}%
                                   {-3.5ex \@plus -1ex \@minus -.2ex}%
                                   {2.3ex \@plus.2ex}%
                                   {\centering\large}}
\makeatother

\makeatletter
\renewcommand\subsection{\@startsection{subsection}{2}{\z@}%
                                     {-3.25ex\@plus -1ex \@minus -.2ex}%
                                     {1.5ex \@plus .2ex}%
                                     {\normalfont\normalsize}}
\makeatother

\makeatletter    

\newcommand{\unnumberedcaption}%
        {\@dblarg{\@unnumberedcaption\@captype}}

\newcommand{\@unnumberedcaption}{}
\long\def\@unnumberedcaption#1[#2]#3{\par
  \addcontentsline{\csname ext@#1\endcsname}{#1}{%
    \protect\numberline{}{\ignorespaces #2}%
    }%
  \begingroup
    \@parboxrestore
    \normalsize
    \@makeunnumberedcaption{\ignorespaces #3}\par
  \endgroup}

\newcommand{\@makeunnumberedcaption}[1]{%
  \vskip\abovecaptionskip
  \sbox\@tempboxa{#1}%
  \ifdim \wd\@tempboxa >\hsize
    #1\par
  \else
    \global \@minipagefalse
    \hbox to\hsize{\hfil\box\@tempboxa\hfil}%
  \fi
  \vskip\belowcaptionskip}

\@ifundefined{abovecaptionskip}{%
  \newlength{\abovecaptionskip}%
  \setlength{\abovecaptionskip}{10pt}%
}{}
\@ifundefined{belowcaptionskip}{%
  \newlength{\belowcaptionskip}%
  \setlength{\belowcaptionskip}{0pt}%
}{}

\makeatother    

\title{Local Polynomial Estimation for Sensitivity Analysis on Models
  With Correlated Inputs}
\date{\empty}

\begin{document}

\renewcommand{\refname}{\begin{center} REFERENCES \end{center}}

\maketitle

\begin{center}
\author{Sebastien DA VEIGA${}^{(1),(2)}$,\ 
  \and Francois WAHL${}^{(1)}$, \ \and Fabrice GAMBOA${}^{(2)}$\\
(1)~: IFP-Lyon, France \\
(2)~: Institut de Mathematiques, Toulouse, France  }
\end{center}

\begin{abstract}

\noindent Sensitivity indices when the inputs of a model are not
independent are estimated by local polynomial techniques. Two original
estimators based on local polynomial smoothers are proposed. Both have
good theoretical properties which are exhibited and also illustrated
through analytical examples. They are used to carry out a sensitivity
analysis on a real case of a kinetic model with correlated
parameters.\\


\end{abstract}

\begin{multicols}{2}

Achieving better knowledge of refining processes usually
requires to build a kinetic model predicting the output
components produced in a unit given the input components
introduced (the ``feed'') and the operating conditions. Such a model
is based on the choice of a reaction mechanism depending on various
parameters (\textit{e.g.} kinetic constants). But the complexity of the
mechanism, the variability of the behavior of catalysts when they are used and the
difficulty of observations and
experiments imply that most often these parameters cannot be inferred from
theoretical considerations and need to be estimated through
practical experiments. This estimation procedure leads to consider
them uncertain and this
uncertainty spreads on the model predictions. This can be highly
problematic in real situations. It is then essential to quantify this
uncertainty and to study the influence of parameters variations on
the model outputs through uncertainty and sensitivity analysis.

During the last decades much effort in mathematical analysis
of physical processes has focused on modeling and reasoning with
uncertainty and sensitivity. Model calibration and validation are
examples where sensitivity and uncertainty analysis have become
essential investigative scientific tools. Roughly speaking,
uncertainty analysis refers to the inherent
variations of a model (\textit{e.g.} a modeled physical process) and is helpful in finding
the relation between some variability or probability distribution on input
parameters and the variability and probability distribution of outputs, while
sensitivity analysis investigates the effects of varying a model
input on the outputs and ascertains how much a model depends on each
or some of its inputs.

Over the years several mathematical and computer-assisted
methods have been developed to carry out global sensitivity analysis
and the reader may refer to the book of \citeasnoun{SCS00} for a wide
and thorough review. Amongst these methods a particular popular class
is the one composed by ``variance-based'' methods which is detailed
below. Let us consider a mathematical model given by 

\begin{equation}
Y=\eta({\bf X})   \label{modele}
\end{equation}
where $\eta : \mathbb{R}^d \rightarrow \mathbb{R}$ is the modeling
function, $Y \in \mathbb{R}$ represents the output or prediction of
the model and ${\bf X}=(X_1,...,X_d)$ is the $d$-dimensional real
vector of the input factors or parameters. The
vector of input parameters is treated as a random vector, which
implies that the output is also a random variable. In variance-based
methods, we are interested in explaining the variance
$\textrm{Var}(Y)$ through the variations of the $X_{i}$,
$i=1,...,d$ and we decompose $\textrm{Var}(Y)$ as follows~:

$$\textrm{Var}(Y)=\textrm{Var}( \E( Y | X_i))+\E( \textrm{Var}(Y | X_i))$$
for all $i=1,...,d$ where $\E( Y | X_i)$ and $\textrm{Var}(Y | X_i)$
are respectively the conditional expectation and variance of $Y$ given
$X_{i}$. The importance of $X_i$ on the variance of
$Y$ is linked to how well $\E( Y |X_i)$ fits $Y$ and can then be measured by
the \textit{first order sensitivity index}

$$S_i=\frac{\textrm{Var}( \E( Y | X_i))}{\textrm{Var}(Y)}$$
also called \textit{correlation ratio}. We can also introduce
sensitivity indices of higher orders to take into account input
interactions. For example, the {\it second
order sensitivity index} for $X_i$ and $X_j$ is

$$S_{ij}=\frac{\textrm{Var}( \E( Y | X_i,X_j))-\textrm{Var}( \E( Y |
X_i))-\textrm{Var}( \E( Y | X_j))}{\textrm{Var}(Y)},$$
and so on for other orders, see \citeasnoun{SCS00} for details.

In the case of independent inputs, two techniques, Sobol
\cite{SOB93} and FAST \cite{CUK73} are the most popular methods for
estimating the $S_i$ indices. Although powerful and computationally efficient, these
methods rely on the assumption of independent inputs which is hard
to hold in many practical cases for kinetic models. Nevertheless, three 
original methods, originated by \citeasnoun{RAT01}, \citeasnoun{JAC04}
and \citeasnoun{OO04}, try to deal with this problem. The
first one sets out to calculate the sensitivity indices by using a
replicated latin hypercube sampling, but this approach requires a large
amount of model evaluations to reach an acceptable precision. The
second one is based on the idea of
building new sensitivity indices which generalize the original ones
by taking into account block of correlations among the inputs. This
method is however useless when many input factors are correlated. The
last approach is that of \citeasnoun{OO04} and rely upon the idea of
approximating the function $\eta$ in model (\ref{modele}) by a
so-called 'kriging' response surface \cite{SWN03} and of computing
analytical expressions of the sensitivity indices based on the results
of the kriging approximation. However appealing and accurate, these
analytical expressions involve multidimensional integrals that are only
tractable when the conditional
densities of the input factors are known and easy to integrate. If
this is not the case
the multidimensional integrals must be approximated numerically, but at
high computational cost. We then propose a new way of estimating
sensitivity indices through an intermediate technique in the sense
that it is based on a sample from the joint density of the inputs and
the output like \citeasnoun{RAT01} but also on a nonparametric
regression model like \citeasnoun{OO04}. This approach does not
require as many model evaluations as \citeasnoun{RAT01} and does not
require to approximate multidimensional integrals
as \citeasnoun{OO04} in the general case.  

In this paper to deal with correlated inputs we consider a
new method based on local polynomial approximations for conditional
moments (see the work of \citeasnoun{FG96} and \citeasnoun{FGH96} on
conditional expectation and the papers of \citeasnoun{FY98} and
\citeasnoun{RWH97} on the conditional variance). Given the form of the
sensitivity indices, local polynomial regression can be used to
estimate them. This approach not only allows to compute a
sensitivity index through an easy black-box procedure but also
reaches a good precision.   

The paper is organized as follows. In Section 1 we
review the methods of \citeasnoun{RAT01},
\citeasnoun{JAC04} and \citeasnoun{OO04} and discuss their merits and
drawbacks. In Section 2 we
propose and study two new estimators for sensitivity indices relying on
local polynomial methods. In Section 3 we present both analytical and
practical examples. In Section 4 we finally give some conclusions and
directions for future research.  

\section{MODELS WITH CORRELATED INPUTS}
When the inputs are independent, Sobol showed that the sum of the
sensitivity indices of all orders is equal to 1, due to an orthogonal
decomposition of the function $\eta$ \cite{SOB93}. Indeed sensitivity
indices naturally arise from this functional
ANOVA decomposition. Nevertheless, when the
inputs are correlated, this property does not hold anymore because such a
decomposition can not be done without taking into account the joint
distribution of the inputs. If one decides to estimate sensitivity
indices under the independence hypothesis although it does not hold,
results and consequently interpretation can be higly misleading, see
the first example of Section 3.1. But we can still consider the initial
ANOVA decomposition and work with the original sensitivity indices
without ignoring the correlation,
and when quantifying the first order sensitivity index of a particular input
factor a part of the sensitivity of all the other input factors
correlated with it is also taken into account. Thus the same
information is considered several times. Interpretation of
sensitivity indices when the inputs are not independent becomes
problematic. However, the input factors being independent or
not, the first-order sensitivity index still points out which factor
(if fixed) will mostly reduce the variance of the output. Thus, if
the goal of the practitioner is to conduct a 'Factors Prioritisation'
\cite{STCR04}, {\it i.e.} identifying the factor that one should fix
to achieve the greatest reduction in the uncertainty of the output,
first-order sensitivity indices remain the measure of importance to
study, see \citeasnoun{STCR04}. Considering that this goal is common
for practitioners, being able to compute first-order sensitivity
indices when the inputs are no longer independent is an interesting
challenge.

Beyond this problem of interpretation, correlation also
makes the computational methods FAST and Sobol unusable as they have
been designed for the independent case. To get over these
difficulties, it is first possible to build 'new'
sensitivity indices that would generalize the original ones and match
their properties, allowing interpretation. This is the idea of
multidimensional sensitivity analysis of Jacques \cite{JAC04} detailed
in the next section. Secondly, \citeasnoun{RAT01} tried to continue on
working with the original sensitivity indices and to compute them as described in
Section 1.3,  even if they do not give clues for interpretation. The
authors generate replicated latin hypercube samples to approximate
conditional densities. Finally, \citeasnoun{OO04} suggest to approach
the function $\eta$ in model (\ref{modele}) by a kriging response surface
which allows to get analytical expressions of sensitivity indices
through multidimensional integrals.

\subsection{Multidimensional Sensitivity Analysis}
To define multidimensional sensitivity indices, \citeasnoun{JAC04}
suggest to split ${\bf X}$ into $p$ vectors ${\bf U}_j,\ j=1,...,p$, 
each of size $k_j$ such that ${\bf U}_j$ is independent from ${\bf
  U}_l$ for $1 \leq j,l \leq p$, $j \neq l$ :

\begin{eqnarray*}
{\bf X}=(X_1,...,X_d) &=&(\underbrace{X_{1},...,X_{k_1}}_{{\bf
U}_{1}},\underbrace{X_{k_1+1},...,X_{k_1+k_2}}_{{\bf
U}_2},...\\
&& ...,\underbrace{X_{k_1+k_2+...+k_{p-1}+1},...,X_{k_1+k_2+...+k_p}}_{{\bf
    U}_p})
\end{eqnarray*}
where $k_1+k_2+....+k_p=d$. For example, if ${\bf
X}=(X_1,X_2,X_3)$ where $X_1$ is independent of $X_2$ and $X_3$ but
$X_2$ and $X_3$ are correlated, we set ${\bf U}_1=X_1$ and ${\bf U}_2=(X_2,X_3)$.

Thus they build \textit{first order multidimensional
sensitivity indices} using the ${\bf U}_j$ vectors :

\begin{eqnarray*}
S_j &=& \frac{\textrm{Var}(\E(Y | {\bf
U}_j))}{\textrm{Var}(Y)}\\
&=&\frac{\textrm{Var}(\E(Y
|X_{k_1+k_2+...+k_{j-1}+1},...,X_{k_1+k_2+...+k_j}))}{\textrm{Var}(Y)}
\end{eqnarray*}
for $j=1,...,p$. Remark that if the inputs are independent,
these sensitivity indices have the same expression as in classical
sensitivity analysis. Finally, it is possible to compute these indices by
Monte-Carlo estimations.

Nevertheless, this method has a main drawback hard
to overcome. If all the inputs are correlated, the ${\bf U}_j$
vectors cannot be defined (except the trivial case ${\bf U}_1={\bf X}$) and
interpretation is not possible. The problem remains the same if too
many inputs are dependent because this situation leads to consider
very few multidimensional indices. Moreover, identifying a set of
correlated variables ${\bf U}_j$ with high sensitivity index does not
allow to point up whether this is due to one particular input of the set
as we cannot differentiate among them. We will illustrate this phenomenon in
the second example of Section 3.1.

\subsection{Correlation-Ratios With Known Conditional Density Functions}
The estimator introduced by \citeasnoun{RAT01} was first discussed in
\citeasnoun{MCK96} and is based on samples from the conditional
density functions of $Y$ given $X_i$, $i=1,...,d$.

Let $({\bf X}^{j})_{j=1,...,n}$ be an i.i.d sample of size $n$
from the distribution of the vector ${\bf X}$. $(X_i^j)_{j=1,...,n}$
is then an i.i.d. sample of size $n$ from the
distribution of the input factor $X_i$. For each realization $X_i^j$
of this sample, let $(Y_i^{jk})_{k=1,...,r}$ be an
i.i.d. sample of size $r$ from the conditional density function of $Y$ given
$X_i=X_i^j$ and define the sample means

$$\overline{Y}_i^j=\frac{1}{r} \sum_{k=1}^r Y_i^{jk} \ \ \
\overline{Y}_i=\frac{1}{n}\sum_{j=1}^n \overline{Y}_i^j.$$
Note that $\overline{Y}_i^j$ and $\frac{1}{r} \sum_{k=1}^{r}(
Y_i^{jk}-\overline{Y}_i^{j})^2$ respectively estimate the conditional
expectation $\E(Y | X_{i}=X_{i}^{j})$ and the conditional variance
$\textrm{Var}(Y|X_{i}=X_{i}^{j})$, while $\overline{Y}_i$ estimates $\E(Y)$.

Using these moments estimators the numerator of the first order
sensitivity index $S_{i}$, $\textrm{Var}(\E(Y|X_{i}))$, can be estimated by the empirical
estimator

$$\frac{1}{n} \sum_{j=1}^{n} (\overline{Y}_i^j -\overline{Y}_i )^{2}.$$
Similarly the denominator of $S_{i}$, $\textrm{Var}(Y)$, is
estimated by

$$\frac{1}{n} \sum_{j=1}^{n} \frac{1}{r} \sum_{k=1}^{r}( Y_i^{jk}-\overline{Y}_i)^2. $$
The estimator of the first order sensitivity index $S_i$ of
the input factor $X_i$, $i=1,...,d$ is then defined as

$$\hat{S}_i=\frac{SSB}{SST}$$
where

$$SSB=r \sum_{j=1}^n (\overline{Y}_i^j-\overline{Y}_i)^2$$
and

$$SST=\sum_{j=1}^n \sum_{k=1}^r ( Y_i^{jk}-\overline{Y}_i)^2.$$
To compute these indices and to generate the samples needed,
Ratto uses two different methods : pure Monte-Carlo sampling and a
single replicated Latin HyperCube (r-LHS) sampling.

It is crucial to note, however, that these two methods require
a huge amount of model evaluations to reach a good precision and can
only be used for cases where model runs have very low computational
cost.  

\subsection{Bayesian Sensitivity Analysis}

The idea of \citeasnoun{OO04} is to see the function
$\eta(\cdot)$ as an unknown smooth function and to formulate a prior
distribution for it. More precisely, it is modeled as the realization of
a Gaussian stationary random field with given mean and covariance
functions. Then, given a set of of values
$y_{i}=\eta(\mathbf{x}_{i})$, we can derive the posterior
distribution of $\eta(\cdot)$ by classical Bayesian considerations.
The prior distribution of $\eta(\mathbf{x})$ is a Gaussian stationary field~:
\begin{equation*}
\eta(\mathbf{x})=\mathbf{h(x)}^{t}\beta + Z(\mathbf{x})
\end{equation*}
conditionally on $\beta$ and $\sigma^{2}$, where $h(\cdot)$ is a
vector of $q$ known regression functions and $Z(\mathbf{X})$ is a Gaussian
stationary random field with zero mean and covariance function
$\sigma^{2}c(\mathbf{x},\mathbf{x}')$. The vector $h(\cdot)$ and the
correlation function $c(\cdot,\cdot)$ are to be chosen in order to
incorporate some information about how the output responds to the
inputs and about the amount of smoothness we require on the output
respectively. We refer the reader to \citeasnoun{SWN03} and to
\citeasnoun{KO01} for a detailed discussion on these choices. The
second stage prior concerns the conjugate prior form for $\beta$ and
$\sigma^{2}$, which is chosen to be a normal inverse gamma
distribution. Now assuming we observe a set $\mathbf{y}$ of $n$
values of $y_{i}=\eta(\mathbf{x}_{i})$,
we can derive that the posterior distribution of $\eta(\cdot)$ given these
data is a Student distribution, see \citeasnoun{OO04} for details.\\
Using this posterior distribution, sensitivity indices can be
computed analytically through multidimensional integrals involving
functions of the observations and the conditional distributions
of the input factors only. The main advantage of
this Bayesian approach is that the model is only evaluated to
calculate the quantities above, {\it i.e.} to 'fit' the response
surface. Once this is done the estimation of sensitivity indices just
involves the conditional distributions of the input factors. When the number
of model runs is fixed, this method clearly reduces the standard errors of the
estimated sensitivity indices obtained by Monte-Carlo methods such as
Sobol (when the input factors are independent) and can still be used
when the input factors are not independent.\\

However, the multidimensional integrals leading to the computation of
the sensitivity indices, if not tractable analytically, need to be
estimated. Let us describe
more particularly one of the estimators proposed in
\citeasnoun{OO04}. We keep the authors notations and denote by $\E^{*}$
the expectations defined with respect to the posterior distribution of
$\eta(\cdot)$. The numerator of the first-order sensitivity index of $Y$
with respect to $X_{1}$ is estimated by
\begin{equation*}
\E^{*}(\V(\E(Y|X_{1})))=\E^{*}(\E(\E(Y|X_{1})^{2}))-\E^{*}(\E(Y)^{2})
\end{equation*}
and one of the quantities involved in the computation of
$\E^{*}(\E(\E(Y|X_{1})^{2}))$ is for example
\begin{eqnarray*}
U_{1}&=&\int_{\R^{d-1}}\int_{\R^{d-1}}\int_{\R}
c(\mathbf{x},\mathbf{x}^{*}) \ 
\textrm{d}F_{-1|1}(\mathbf{x}_{-1}|x_{1})\\
&& \textrm{d}F_{-1|1}(\mathbf{x'}_{-1}|x_{1})\ 
\textrm{d}F_{1}(x_{1})
\end{eqnarray*}
where $F_{-1|1}$ is the marginal distribution of $\mathbf{X}_{-1}$ (subvector
of $\mathbf{X}$ containing all elements except $X_{1}$) given $X_{1}$,
$F_{1}$ is the marginal distribution of $X_{1}$ and $\mathbf{x}^{*}$ denotes
the vector with elements made up of $\mathbf{x}_{1}$ and
$\mathbf{x}'_{-1}$ in the same way as $\mathbf{x}$ is composed of
$\mathbf{x}_{1}$ and $\mathbf{x}_{-1}$. If the
conditional distribution $F_{-1|1}$ is not analytically known, we
first need to estimate it with a sample of the joint distribution $F$.
Many methods have been developed to do so, let us just mention for example kernel techniques. But in general in high dimension the data is very sparsely distributed and it is difficult to get an accurate approximation of conditional distributions since the so-called curse of dimensionality arises. For instance the best possible MSE rate with kernel techniques is $n^{-4/(4+d)}$ which becomes worse as $d$ gets larger.\\
Moreover, even if we could get a good approximation of $F_{{-1|1}}$,
still remains the problem of evaluating the multidimensional
integrals. Indeed the dimensionality of these integrals can reach
$2d-1$ as in the expression of $U_{1}$ above. Since these integrals can not in general be separated into unidimensional integrals,
approximating them with a sufficent accuracy is not an obvious
mathematical problem. Deterministic schemes can not reasonably be
considered, and with Monte-Carlo or quasi Monte-Carlo sampling \cite{OWE04} thousands (or millions) of draws are required to get a reasonable accuracy.\\  

With unknown densities, even if conceptually, sampling rather than
analytical integration in the Oakley and O'Hagan approach seems
reasonable, the results could be highly affected by the curse of
dimensionality. Let us mention that Pr. O'Hagan has public domain software carrying out this analysis. However it does not yet allow to consider dependent inputs.

\section{NEW ESTIMATION METHODOLOGY}
Our approach is to estimate the
conditional moments $\E(Y | X_{i}=X_{i}^{j})$ and 
$\textrm{Var}(Y|X_{i}=X_{i}^{j})$ with an intermediate method between
the one of \citeasnoun{RAT01} and \citeasnoun{OO04}. We first 
use a sample $(X_{i},Y_{i})$ to estimate the conditional moments with
nonparametric tools (provided they are smooth functions of the input
factors). Then, we compute sensitivity indices by using another sample
of the input factors only (and thus no more model runs are
needed). While \citeasnoun{OO04} approximate the function
$\eta(\mathbf{X})$ in $\R^{d}$, we approximate it marginally, {\it i.e.}
we approximate the conditional
expectations $\E(\eta(\mathbf{X})|X_{i})$ in $\R$. This approach allows
to overcome the multidimensional integration problem of the Bayesian
sensitivity analysis.\\  

To simplify the notations, until Section
2.4 $(X,Y)$ will stand for a bivariate random vector (\textit{i.e.} $X$ is
unidimensional). As the variance may be decomposed as
$\textrm{Var}(Y)=\textrm{Var}( \E( Y | X))+\E( \textrm{Var}(Y |
X))$, the index we wish to estimate can be written 
\begin{equation} 
S=\frac{\textrm{Var}( \E( Y | X))}{\textrm{Var}(Y)} \ \ \textrm{or}
\ \ S=1-\frac{\E( \textrm{Var}(Y | X))}{\textrm{Var}(Y)}.
\label{SI}
\end{equation}
These expressions clearly give two ways of estimating $S$~: the issue
is to be able to estimate $\textrm{Var}( \E( Y | X))$ or alternatively
$\E( \textrm{Var}(Y | X))$, obviously by estimating first the
conditional moments $ \E( Y | X=x)$ and $\textrm{Var}(Y | X=x)$ ($x\in\R$). In both
cases the denominator term $\textrm{Var}(Y)$ can be easily
estimated. To approximate the conditional moments,
we propose to use local polynomial regression. This highly
statistical efficient tool is easy to apprehend as it is close to the
weighted least-squares approach in regression problems. Only basic results
will be presented here, for a detailed
picture of the subject the interested reader is referred to
\citeasnoun{FG96}.\\

\subsection{Formulation of the Estimators}
Let $(X_i,Y_i)_{i=1,...,n}$ be a two-dimensional i.i.d. sample of a
real random vector $(X,Y)$. Assuming that $X$ and $Y$ are
square integrable we may write an heteroskedastic
regression model of $Y_i$ on $X_i$, exhibiting the conditional expectation
and variance, as
$$Y_i=m(X_i)+\sigma(X_i)\epsilon_i, \ \ \ \ \ i=1,\ldots,n$$
where $m(x)=\E(Y|X=x)$ and $\sigma^2(x)=\textrm{Var}(Y|X=x)$
($x\in\R$) are the
conditional moments and the errors $\epsilon_1,\ldots,\epsilon_{n}$
are independent random variables satisfying $\E(\epsilon_i|X_i)=0$ and
$\textrm{Var}(\epsilon_i|X_i)=1$. Usually
$\epsilon_{i}$ and $X_{i}$ are assumed to be independent although this is
not the case in our work. Note that results for correlated
errors have been recently developed (\citeasnoun{FF02} for the autoregressive case for
example). Local polynomial
fitting consists in approximating {\it locally} the regression function $m$
by a $p$-th order polynomial 
$$m(z)\approx \sum_{j=0}^{p}\beta_{j}(z-x)^{j}$$
for $z$ in a neighborhood of $x$. This polynomial is then fitted to
the observations $(X_{i}, Y_{i})$ by solving the weighted least-squares problem
\begin{equation}
\min_{\beta}\sum_{i=1}^n \Bigl( Y_i-\sum_{j=0}^{p}\beta_{j}(X_i-x)^{j}\Bigr)^2
K_{1}\Bigl( \frac{X_i-x}{h_1}\Bigr)   \label{reg}
\end{equation}
where $K_{1}(.)$ denotes a {\it kernel} function and $h_1$ is a
{\it smoothing} parameter (or {\it bandwidth}). In this case, 
if  $\hat{\beta}(x)=(\hat{\beta}_{0}(x),...,\hat{\beta}_{p}(x))^{T}$ denotes the
minimizer of (\ref{reg}) we have
$$\hat{m}(x)=\hat{\beta}_{0}(x),$$
while the $\nu$-th derivative
of $m(x)$ is estimated via the relation 
$$\hat{\beta}_{\nu}(x)=\frac{\hat{m}^{(\nu)}(x)}{\nu !},$$
see \citeasnoun{FG96} for more details. As it will be discussed later,
the smoothing parameter $h_{1}$ is chosen to balance bias and
variance of the estimator. Finally, remark that the particular
case $p=0$ (constant fit) leads to the well-known {\it Nadaraya-Watson} estimator
$\hat{m}_{NW}(x)$ of the conditional expectation, given explicitly by 
$$\hat{m}_{NW}(x)=\frac{\displaystyle{\sum_{i=1}^{n}
  Y_{i}K\Bigl(\frac{X_{i}-x}{h}\Bigr)}}{\displaystyle{\sum_{i=1}^{n}K\Bigl(\frac{X_{i}-x}{h}\Bigr)}},$$
see \citeasnoun{WJ94}.

Estimation of the conditional variance is
less straightforward. If the regression function $m$ was
known, the problem of estimating $\sigma^2(.)$ would be regarded as a
local polynomial regression of $r_i^{2}$ on $X_i$ with
$r_i^{2}=(Y_i-m(X_i))^2$, as $\E(r^{2}|X=x)=\sigma^2(x)$ with
$r^{2}=(Y-m(X))^2$. But in practice, $m$ is unknown. A natural approach
is to substitute $m(.)$ by its estimate $\hat{m}(.)$ defined as above
and to get the {\it the residual-based estimator} $\hat{\sigma}^2(x)$
by solving as previously the weighted least-squares problem 
\begin{equation}
\min_{\gamma}\sum_{i=1}^n \Bigl( \hat{r}_i^{2}-\sum_{j=1}^{q}\gamma_{j}(X_i-x)^{j}\Bigr)^2
K_{2}\Bigl( \frac{X_i-x}{h_2}\Bigr)     \label{reg2}
\end{equation}
where $\hat{r}_i^{2}=(Y_i-\hat{m}(X_i))^2$, $K_{2}(.)$ is a kernel and $h_2$ a
smoothing parameter. Note that the kernel $K_{2}(.)$ is not necessarily
chosen to be equal to the kernel $K_{1}(.)$. Then 
$$\hat{\sigma}^2(x)=\hat{\gamma}_{0}(x)$$
where $\hat{\gamma}(x)=(\hat{\gamma}_{0}(x),...,\hat{\gamma}_{q}(x))$ is the
minimizer of (\ref{reg2}). As previously, the
smoothing parameter $h_{2}$ has to be chosen to balance bias and
variance of the estimator, see \citeasnoun{FY98}.

Going back over the equalities in (\ref{SI}), the last step is to
estimate the quantities $\textrm{Var}( \E( Y | X))$ and $\E( \textrm{Var}(Y | X))$ by using
the local polynomial estimators for the conditional moments defined
right above. To do this let us assume we have another i.i.d. sample
$(\tilde{X}_{j})_{j=1,\ldots,n'}$ with same distribution as $X$. If
the functions $m(.)$ and $\sigma^{2}(.)$
were known, we could estimate $\textrm{Var}( \E( Y |
X))=\textrm{Var}(m(X))$ and $\E( \textrm{Var}(Y |
X))=\E(\sigma^{2}(X))$ with the classical empirical moments
$$\frac{1}{n'-1}
\sum_{j=1}^{n'} \left( m(\tilde{X}_{j})-\bar{m} \right)^{2} \ \
\textrm{and} \ \  \frac{1}{n'}
\sum_{j=1}^{n'} \sigma^{2}(\tilde{X}_{j}) $$
where $\displaystyle{\bar{m}=\frac{1}{n'} \sum_{j=1}^{n'}
  m(\tilde{X}_{j})}$. As $m(.)$ and $\sigma^2(.)$ are unknown, the
main idea is to replace them by their local polynomial estimators
which leads to consider
$$  \hat{T}_{1}=\frac{1}{n'-1}
\sum_{j=1}^{n'} \left( \hat{m}(\tilde{X}_{j})-\hat{\bar{m}}
\right)^{2} \ \ \textrm{and} \ \  \hat{T}_{2}=\frac{1}{n'}
\sum_{j=1}^{n'} \hat{\sigma}^{2}(\tilde{X}_{j})$$
where $\displaystyle{\hat{\bar{m}}=\frac{1}{n'} \sum_{j=1}^{n'}
  \hat{m}(\tilde{X}_{j})}$ and $\hat{m}(.)$ and $\hat{\sigma}^{2}(.)$
are the local polynomial estimators of $m(.)$ and $\sigma^{2}(.)$
introduced above. It is important to note that we need two samples,
the first one $(X_{i},Y_{i})_{i=1,\ldots,n}$ to compute $\hat{m}(.)$ and
$\hat{\sigma}^{2}(.)$ and the second one
$(\tilde{X}_{j})_{j=1,\ldots,n'}$ to finally compute the empirical
estimators $\hat{T}_{1}$ and $\hat{T}_{2}$.

\subsection{Bandwidth and Orders Selection}
The selection of the smoothing parameters $h_{1}$ and $h_{2}$ and to a
lesser extent of the
polynomials orders $p$ and $q$ can be crucial to get the least mean
squared error (MSE) of the estimators $\hat{T}_{1}$ and
$\hat{T}_{2}$. Classically the MSE consists of a bias term plus a variance term
and so is minimized by finding a compromise between bias and
variance.

Concerning this choice, the reader is referred to \citeasnoun{FGH96}, 
\citeasnoun{FY98} or \citeasnoun{R97}. Most of the methods suggested
by these authors rely upon
asymptotic arguments and their efficiency for finite sample cases
is not clear. In practice cross-validation
methods can be used for the finite sample case \cite{JMS96},
but in the examples of Section 3 we will use the empirical-bias bandwidth selector (EBBS) of Ruppert
which appears to be efficient on simulated data. EBBS is based on
estimating the MSE empirically and not with an asymptotic expression.
The choice of the polynomials orders is more
subjective. Concerning the estimation of the conditional expectation,
\citeasnoun{FG96} recommend to
use a $\nu+1$ or $\nu+3$th-order polynomial to estimate the
$\nu$th-derivative of $m(x)$, following theoretical considerations on
the asymptotic bias of $\hat{m}(x)$ on the boundary. We would then be
lead to take $p=1$ or $p=3$ to
estimate the $0$th-derivative $m(x)$. But \citeasnoun{RWC03} suggest
that this conclusion should be balanced by simulation studies and
stress that $p=2$ often outperforms $p=1$ and $p=3$. The only
common conclusion is that
local linear regression ($p=1$) is usually superior to kernel
regression (Nadaraya-Watson estimator obtained with $p=0$). This
is the reason why we will only consider and study local linear
regression for $m(x)$ in the next theoretical and practical
sections. The choice is still difficult when estimating the
conditional variance as we have to choose $p$ and $q$
simultaneously. One more time, the authors are not unanimous~:
\citeasnoun{FY98} recommend the case $p=1,q=1$ whereas
\citeasnoun{RWH97} suggest $p=2,q=1$ or $p=3,q=1$. However on the simulations
we have carried out, the choice of $p=1, q=1$ is adequate and
satisfactory in terms of precision. This is the reason why we have
decided to consider only the case $p=1,q=1$ for both theoretical and
practical results.     

\subsection{Theoretical Properties of the Estimators}
The properties of $\hat{T}_{1}$ and $\hat{T}_{2}$ strongly depend on
the asymptotic results on the bias and variance of the local linear
estimators $\hat{m}(.)$ and $\hat{\sigma}^{2}(.)$. We only give here
two main results, all assumptions $(A_{0}, ..., A_{4}, B_{0}, ...,
B_{4}, C_{0})$ and proofs are
given in appendix for readability. $\E_{\mathbb{X}}$ and
$\textrm{Var}_{\mathbb{X}}$ stand for the conditional expectation and
variance given the predictors $\mathbb{X}=(X_{1},...,X_{n})$. The
expression $o_{P}(\varphi(h))$ is equal to $ \varphi(h)o_{P}(1)$ for a
given function $\varphi$. Here $o_{P}(1)$
is the standard notation for a sequence of random variables that
converges to zero in probability.   

\begin{theoreme}
Under assumptions (A0)-(A4) and (C0), the estimator $\hat{T}_{1}$ is
asymptotically unbiased. More precisely

$$\E_{\mathbb{X}}(\hat{T}_{1})=\textrm{Var}(\E(Y|X))+M_{1}h_{1}^{2}+\frac{M_{2}}{nh_{1}}+o_{P}(h_{1}^{2}).$$
where $M_{1}$ and $M_{2}$ are constants given in appendix.

\end{theoreme}

{\it Remark 1.} It would be interesting to calculate the variance of
this estimator, but it would require the expressions of the third and
fourth moments of the local linear
estimator $\hat{m}(.)$ (see the appendix). This is not an obvious problem and
 to the best of our knowledge it has not been addressed in the
literature. It is beyond the scope of the present paper but it is an
interesting problem for future research. Nevertheless, the variance
can be estimated on practical cases through bootstrap
methods for example \cite{ET94}.

\begin{theoreme}
Under assumptions (B0)-(B4) and (C0), the estimator $\hat{T}_{2}$ is
consistent. More precisely

$$\E_{\mathbb{X}}(\hat{T}_{2})=\E(\textrm{Var}(Y|X))+V_{1}h_{2}^{2} + o_{P}(h_{1}^{2}+h_{2}^{2})$$
and

\begin{eqnarray*}
\textrm{Var}_{\mathbb{X}}(\hat{T}_{2})&=&\frac{1}{n'} \left\{
  \E(\textrm{Var}(Y|X)^{2}) + V_{2} h_{2}^{2}+ V_{3}h_{1}^{2}
  +\frac{V_{4}}{nh_{2}} \right.\\
&&  + \left. o_{P}\left(h_{1}^{2}+h_{2}^{2}+ \frac{1}{\sqrt{nh_{2}}}\right)
\right\}
\end{eqnarray*}

where $V_{1}$, $V_{2}$, $V_{3}$ and $V_{4}$ are constants given in
appendix.

\end{theoreme} 
 
\subsection{Application to Sensitivity Analysis}
Let us come back to the model (\ref{modele}), where ${\bf X}$ is
multidimensional. The goal is to get an estimate of $S_{i}$
 for $i=1,\ldots,d$ by using one of the two estimators $\hat{T}_{1}$
 and $\hat{T}_{2}$. We need two samples to compute each of them, {\it i.e.} a
 sample $(X^{k}_{i},Y^{k})_{k=1,\ldots,n}$ to estimate $\hat{m}(.)$
 and $\hat{\sigma}^{2}(.)$ and a sample
 $(\tilde{X}^{l}_{i})_{l=1,\ldots,n'}$ to get $\hat{T}_{1}$ and
 $\hat{T}_{2}$ where $(X_{i}^{k})_{k=1,\ldots,n}$ and
 $(\tilde{X}^{l}_{i})_{l=1,\ldots,n'}$ are samples from the joint
 distribution of the $d$-dimensional input factors
 $\mathbf{X}=(X_{i})_{i=1,\ldots,d}$ and $(Y^{k})_{k=1,\ldots,n}$ a
 sample of the output $Y$. Note that the model is run just for the
 first sample and not for the second one. Three situations can arise~:
\begin{itemize}
\item[1.] Sampling from the joint distribution of
$\mathbf{X}$ has low computational cost and running the model to compute
$(Y^{k})_{k=1,\ldots,n}$ is cheap. This is the ideal
situation. Indeed in this case the two samples
$(X^{k}_{i},Y^{k})_{k=1,\ldots,n}$ and
$(\tilde{X}^{l}_{i})_{l=1,\ldots,n'}$ can be generated independently
and be as large as required\,;
\item[2.] Sampling from the joint distribution of
$\mathbf{X}$ has low computational cost but model evaluations have
not. In this case (also pointed out by \citeasnoun{OO04}) a
moderate-sized sample $(X^{k}_{i},Y^{k})_{k=1,\ldots,n}$ is used in
order to fit the conditional moments. However to compute $\hat{T}_{1}$ and
 $\hat{T}_{2}$ we can then use a sample
 $(\tilde{X}^{l}_{i})_{l=1,\ldots,n'}$ of large size\,;
\item[3.] Sampling from the joint distribution of
$\mathbf{X}$ has high computational cost. This case can arise in
practice for example when the input factors are obtained through a procedure based
on experimental data and optimization routines. We then have an
initial sample $(\mathbf{X}^{j})_{j=1,\ldots,N}$ of limited size $N$
that we wish to use for the two steps of the estimation. The first
idea is to split it and to use the first part to get the sample
$(X^{k}_{i},Y^{k})_{k=1,\ldots,n}$ and the second one to get
$(\tilde{X}^{l}_{i})_{l=1,\ldots,n'}$. The drawback of this method
clearly arises if $N$ is very small. Another way to tackle the problem
is to use the well-known leave-one-out idea procedure which gives
better approximation than data splitting.\\
As suggested by the Associate Editor another possible method could be
to use the sample of size $N$ to estimate the
conditional moments and to estimate also the marginal densities of
each input using for instance a nonparametric density estimator. One
could then use these density estimates to get the sample
$(\tilde{X}^{l}_{i})_{l=1,\ldots,n'}$. The clear disadvantage of this
procedure is that it may bias the final estimators. Some simulation
runs not reported here for lack of space show that such a procedure
leads to less efficient estimates probably due to the large bias
produced by nonparametric methods.
\end{itemize}

The last situation obviously leads to the less accurate approximations
of first-order sensitivity indices. However in general, litterature and
results on sensitivity analysis assume that, if not analytically
known, the joint distribution of the input factors can at least be
generated at low computational cost. This is the reason why we will
only describe here the procedure for estimating first-order
sensitivity indices in case $1$ or $2$. We now assume that we have two samples
$(X^{k}_{i})_{k=1,\ldots,n}$ and $(\tilde{X}^{l}_{i})_{l=1,\ldots,n'}$
obtained by one of the methods described right above.

\noindent  The estimation procedure for
$\displaystyle{S_i=\frac{\textrm{Var}(\E(Y|X_i))}{\textrm{Var}(Y)}}$
is the following :\\

Step 1 : Compute the output sample  $(Y^k)_{k=1,...,n}$ by
running the model at $({\bf X}^k)_{k=1,...,n}$ \\

Step 2 : Compute $\hat{\sigma}_{Y}^2$, the classical unbiased estimator
of the variance $\textrm{Var}(Y)$ 

$$\hat{\sigma}_{Y}^2=\frac{1}{n-1} \sum_{k=1}^n \left(Y^k -
  \bar{Y}\right)^{2} $$

Step 3 : Use the sample $(X^k_i,Y^{k})_{k=1,...,n}$ to obtain
$\hat{m}(\tilde{X}^{l}_{i})$ for $l=1,\ldots,n'$ and
$\hat{m}(X^{k}_{i})$ for $k=1,\ldots,n$ using the smoothing
parameter $h_{1}$ given by EBBS \\

Step 4 : Compute squared residuals
$\hat{r}_{k}=(Y^{k}-\hat{m}(X^{k}_{i}))^{2}$ for $k=1,\ldots,n$ and
apply the smoothing parameter $h_{2}$ obtained by EBBS to
compute $\hat{\sigma}^{2}(\tilde{X}^{l}_{i})$ for
$l=1,\ldots,n'$ \\

Step 5 : Compute $\hat{T}_{1}$ with $\hat{m}(\tilde{X}^{l}_{i})$ for
$l=1,\ldots,n'$ from Step 3 and compute $\hat{T}_{2}$ with
$\hat{\sigma}^{2}(\tilde{X}^{l}_{i})$ for $l=1,\ldots,n'$ from Step 4\\ 

Step 6 : The estimates of $S_i$ are then 

\begin{equation}
  \hat{S}_i^{(1)}=\frac{\hat{T}_{1}}{\hat{\sigma}_{Y}^2} \ \
  \textrm{and} \ \ \hat{S}_i^{(2)}=1-\frac{\hat{T}_{2}}{\hat{\sigma}_{Y}^2}.   \label{S1}
\end{equation}

To obtain all the first-order sensitivity indices, repeat
the procedure from Step 3 to Step 6 for $i=1,...,d$.

{\it Remark 2.} Given the theoretical properties of $\hat{T}_{1}$ and
$\hat{T_{2}}$ and more precisely their non-parametric convergence
rate, we can also expect a nonparametric convergence rate for
$\hat{S}^{(1)}$ and $\hat{S}^{(2)}$.

{\it Remark 3.} In practice, our simulations show that $n$ of the
order of $100$ and $n'$ around $2000$ are enough for accurate
estimation of the sensitivity indices.

\section{EXAMPLES}
In all the following examples we use the two estimators
$\hat{S}^{(1)}$ and $\hat{S}^{(2)}$ defined in (\ref{S1}). As
mentioned in Section 2.2, the conditional expectation is estimated
here with local linear regression ($p=1$) and the conditional variance
with $p=1$ and $q=1$, the bandwidths being selected by the
estimated-bias method of \citeasnoun{R97}.\\

\subsection{Analytical Examples}
In this section, we carry out two different comparisons in order to study
our two estimators from a numerical point of view. The first model has
been chosen to underline their precision in correlated cases when FAST
and Sobol methods are no longer efficient and when Jacques' approach
for multidimensional sensitivity analysis is limited. We also show how
interpretation with sensitivity indices obtained by neglecting
correlation can be false. The second one is an example illustrating
the performance of our estimators with respect to the method of Oakley
and O'Hagan in a two-dimensional setting.

In the first analytical example, we study the model
$$Y=X_1 + X_2 + X_3$$
where $(X_1,X_2,X_3)$ is a three-dimensional normal vector with mean
${\bf 0}$ and covariance matrix
$$\Gamma=\left[ \begin{array}{ccc} 1 & 0 & 0\\
0 & 1 & \rho \sigma\\ 0 & \rho \sigma & \sigma^2 \end{array} \right]$$
where $\rho$ is the correlation of $X_2$ and $X_3$ and $\sigma>0$ is the
standard deviation of $X_3$. The first order sensitivity indices can
be evaluated analytically~: 
$$S_1=\frac{1}{2+\sigma^2+2\rho \sigma}$$
$$S_2=\frac{ (1+\rho \sigma)^2}{2+\sigma^2+2\rho \sigma}$$
$$S_3=\frac{(\sigma+\rho)^2}{2+\sigma^2+2\rho \sigma}$$
The first crucial remark to be done in this case is that we must take
into account correlations to estimate sensitivity indices if we want a
serious investigation of this model. Indeed, let us consider the case
where $\sigma=1.2$ and $\rho=-0.8$. We then have
$$S_1=0.6579, \ \ \ S_2=0.0011, \ \ \ S_3=0.1053,$$
indicating that $X_{1}$ should be the input to be fixed to reach the
higher variance reduction on $Y$. But if one neglects the correlation,
by computing for instance these indices with the FAST method, {\it
  i.e.} working with a three-dimensional normal vector with mean
$\mathbf{0}$ and covariance matrix $I$ instead of $\Gamma$, one would
estimate 
$$S^{0}_1=0.2907, \ \ \ S^{0}_2=0.2907, \ \ \ S^{0}_3=0.4186$$
where $S^{0}$ stands for the sensitivity indices when $\rho=0$. These
results indicate that $X_{3}$ should be fixed to mostly reduce the
variance of $Y$, which is absolutely wrong as the calculations above
have shown. This simple example highlights the danger of neglecting
the correlations between the inputs and the importance to take them
into consideration when computing sensitivity indices.\\  

\noindent Otherwise, applying Jacques' idea to $X_1$ and the couple $(X_2,X_3)$,
we also get the expression of the first order multidimensional
sensitivity index
$$S_{\{2,3\}}=\frac{1+\sigma^2+2\rho \sigma}{2+\sigma^2+2\rho \sigma}$$
Choosing $\rho=-0.2$ and $\sigma=0.4$, we have
$$S_1=S_{\{2,3\}}=0.5, \ \ \ S_2=0.4232, \ \ \ S_3=0.02$$
If we interpret these indices as suggested by Jacques'
multidimensional sensitivity analysis, the only conclusion we can give
is that the couple $(X_2,X_3)$ has the same importance as
$X_1$. Indeed $S_{\{2,3\}}=S_1$. But actually the high value of $S_{\{2,3\}}$ comes
from $X_{2}$ as shown by the exact calculations above, which implies
that the information on $S_{\{2,3\}}$ alone is not sufficient. But
with our method, we can estimate all the first order sensitivity
indices~:
$$ \hat{S}_1^{(1)}=0.4895, \ \ \ \hat{S}_2^{(1)}=0.4250, \ \
\ \hat{S}_3^{(1)}=0.0234$$
$$ \hat{S}_1^{(2)}=0.5081, \ \ \ \hat{S}_2^{(2)}=0.4368, \ \
\ \hat{S}_3^{(2)}=0.0361$$
for an average upon 100 simulations with $n=50$ and $n'=1000$. We
display in Figure 1 the boxplots corresponding to the
distribution of the sensitivity indices on these 100 simulations with
the estimator $\hat{T}_2$. Because of the mathematical complexity
mentioned before for the computation of the variance of $\hat{T}_{1}$,
we are not able to recommend one estimator over the other one from a
theoretical point of view. But in practice, we have observed that the
variance of $\hat{T}_{2}$ is at least comparable to the variance of
$\hat{T}_{1}$, and sometimes lower. Nervertheless, the computation of
$\hat{T}_{2}$ is more difficult as illustrated in Section 2.4.

\begin{figurehere}
      \includegraphics[width=0.8\linewidth]{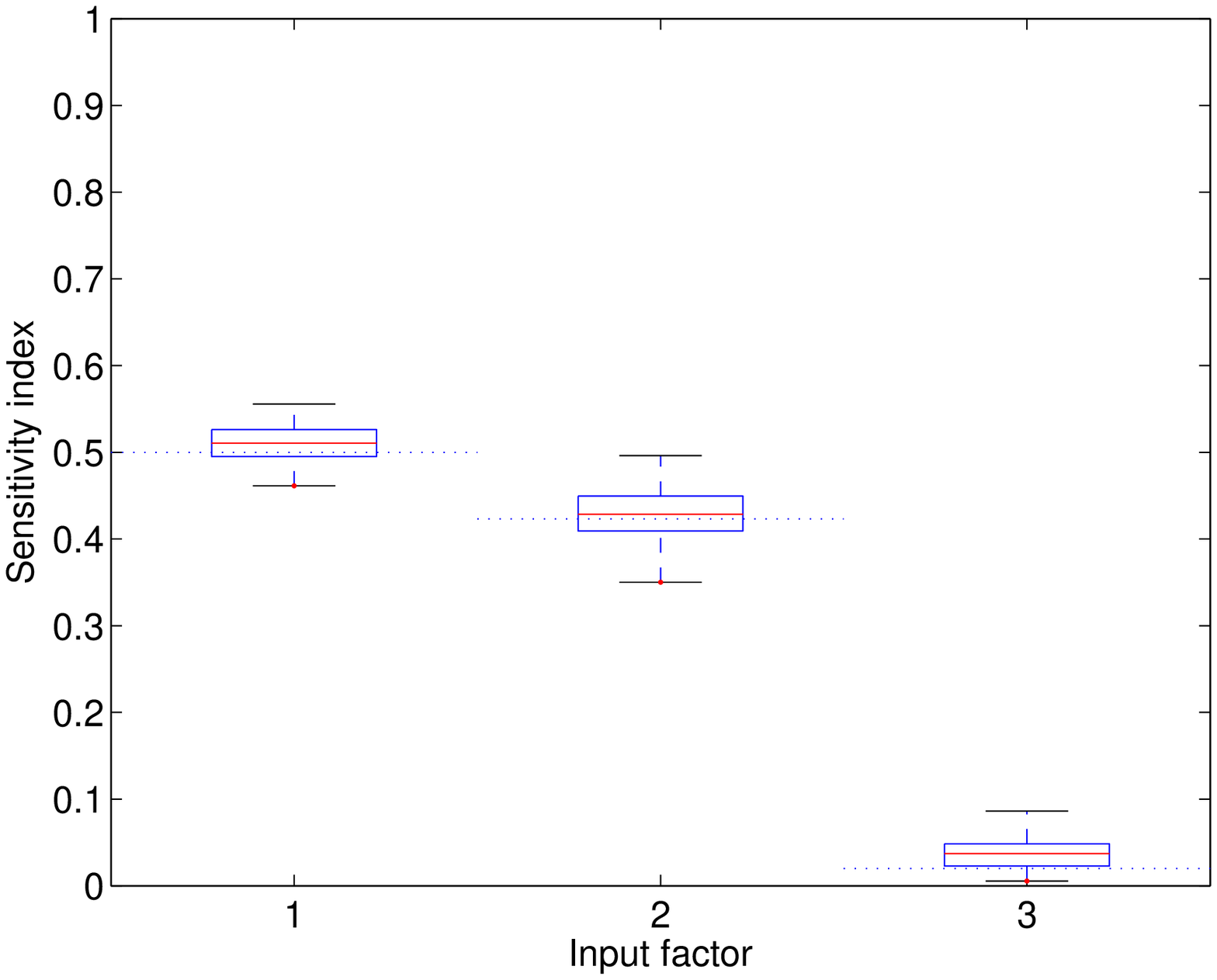}
        \unnumberedcaption{{\small \it Figure 1. Boxplot of the estimated sensitivity indices ($\hat{S}^{(2)}$)
            for the three-factor additive model, 100 simulations. Dot
            lines are the true values}.\\}
\end{figurehere}
Computing $S_2$ and $S_3$ with our method, even
if both of them take into account correlations, allows to confirm the
expected result : all the variability comes from $X_2$, and not from
$X_3$. This simple example then brings out the limitation of the
multidimensional approach.\\

In the second analytical example we consider the model 
\begin{eqnarray*}
Y&=&0.2\exp(X_{1}-3)+2.2|X_{2}|+1.3X_{2}^6-2X_{2}^2\\
&&-0.5X_{2}^4-0.5X_{1}^4+2.5X_{1}^2+0.7X_{1}^3\\
&&+\frac{3}{(8X_{1}-2)^2+(5X_{2}-3)^2+1}+\sin(5X_{1})\cos(3X_{1}^2)
\end{eqnarray*}
where $X_{1}$ and $X_{2}$ are independent random variables uniformly
distributed on $[-1,1]$. Such a model is routinely used at Institut Francais
du Petrole to compare different response surface methodologies as it
presents a peak and valleys. The function is plotted in Figure 2.

\begin{figurehere}
      \includegraphics[width=0.8\linewidth]{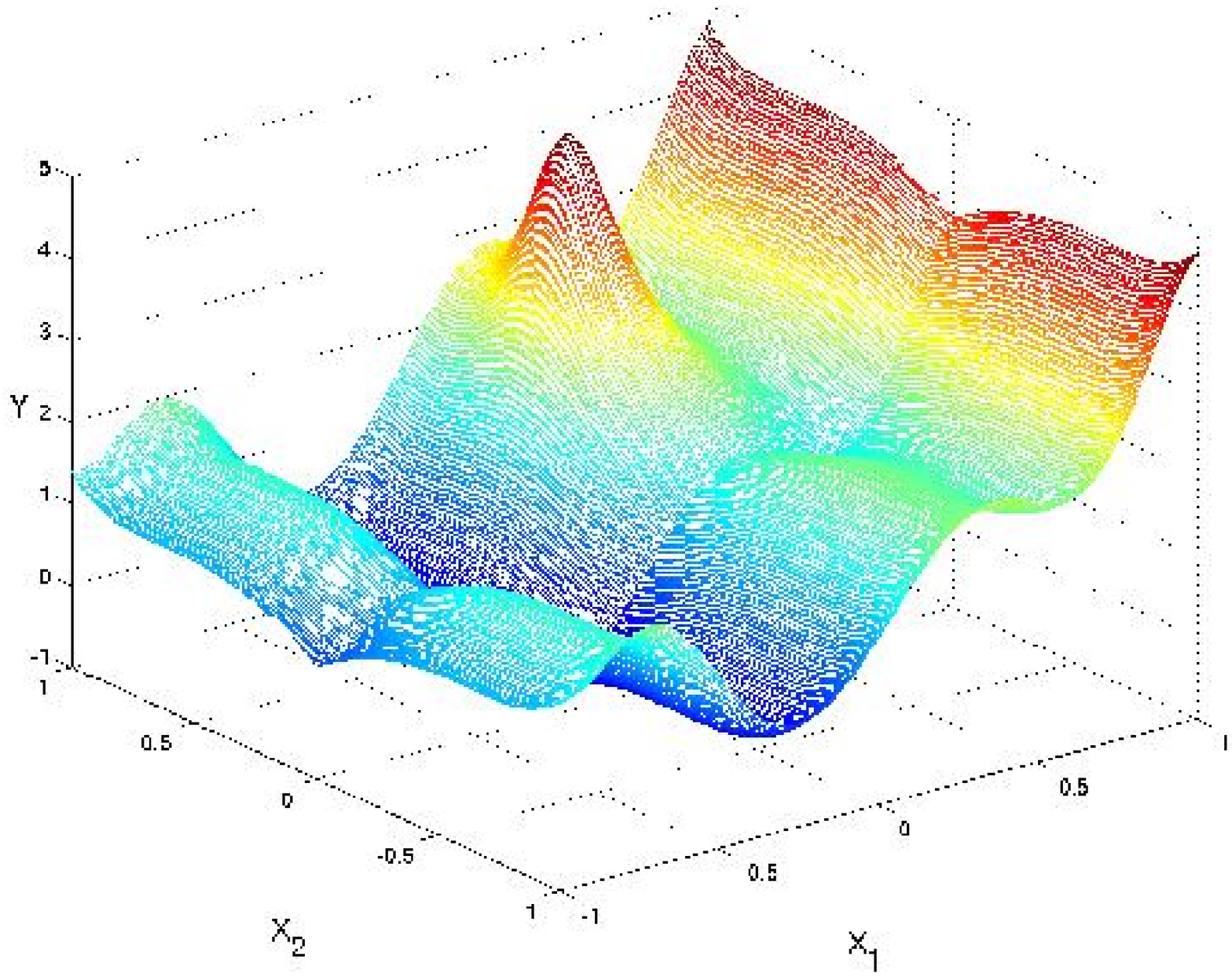}
        \unnumberedcaption{{\small \it Figure 2. Function proposed in
            model 2 .}}
\end{figurehere} 

In this case, the sensitivity indices are
\begin{equation*}
S_{1}=0.9375 \ \ \textrm{and} \ \ S_{2}=0.0625
\end{equation*}

We considered a $6\times 6$ regular grid on $[0,1]^{2}$ and used it to
estimate the posterior distribution in the method of Oakley and
O'Hagan and to estimate the conditional moments in our method. Then,
we calculated analytically the multidimensional integrals in the
Bayesian approach while using a sample of size $5000$ to compute
$\hat{S}^{(2)}_{i}$ for $i=1,2$. The Bayesian
approach leads to  
\begin{equation*}
\hat{S}_{1}=0.9038 \ \ \textrm{and} \ \ \hat{S}_{2}=0.0961
\end{equation*}
while the local polynomial technique gives
\begin{equation*}
\hat{S}^{(2)}_{1}=0.9127 \ \ \textrm{and} \ \ \hat{S}^{(2)}_{2}=0.0452.
\end{equation*}
We can see on this example that the results obtained with both methods
are comparable. However on this simple case the multidimensional
integrals were analytically computed, which could not be the case in a
non-independent setting. If not, a numerical integration, if feasible,
would lead to less accurate approximations as discussed in Section
1.3.

\subsection{Practical Example from Chemical Field : Isomerization of
  the Normal Butane}
The isomerization of the normal butane, {\it i.e.} molecules with four carbon
atoms, is a chemical process aiming at transforming normal butane (nC4)
into iso-butane (iC4) in order to obtain a higher octane number,
favored by iC. A simplified reaction mechanism has been used :
$$\textrm{nC4} \longleftrightarrow \textrm{iC4} \ \ \ \ \ \ \ \ \ \ \ \ \ \ \ \ (1)$$
$$2 \ \textrm{iC4} \longrightarrow \textrm{C3+C5} \ \ \ \ \ \ \ \ \ \ (3)$$
$$\textrm{nC4}+\textrm{iC4} \longrightarrow \textrm{C3+C5} \ \ \ \ (4)$$
where reaction $(1)$ is the main reversible reaction converting the
normal butane into
iso-butane. Reactions $(3)$ and $(4)$ are secondary and irreversible
reactions which produce propane (C3) and a lump of normal and
iso-pentane (C5), paraffins with three and five
carbon atoms. The model linked
to this process can be written as
$${\bf Y}=\eta({\bf c},\boldsymbol{\theta})$$
where \\

- ${\bf Y}$ is the 3-dimensional result vector (mole fractions of the
components nC4, iC4, C3 and C5\,; note that their sum is 1),\\

- ${\bf c}$ is the vector containing the operating conditions
(pressure, temperature,...) and the mole fraction of the input
components (nC4 and iC4, this is called the {\it feed}),\\

- $\boldsymbol{\theta}=(\theta^{i})_{i=1,...,8}$ is the 8-dimensional random vector of the
parameters of the reactions (pre-exponential factors, activation
energies, adsorption constants,...),\\

- $f$ is the function modeling the chemical reactor in which the
reaction takes place. It is evaluated through the resolution of an ordinary
differential equations system which can not be analytically solved and is
calculated numerically.\\

The first step here is to get the distribution of $\boldsymbol{\theta}$ which is unknown. However, it is possible to use the experience and the knowledge of chemical engineers to suggest a reasonable approximation of this distribution. Classically, we assume that $\boldsymbol{\theta}$ has a multivariate Gaussian distribution with mean zero (once the parameters are centered). Concerning the correlation matrix, it is built with experts and with the help of bootstrap simulations and is given by :
{\footnotesize
\begin{equation*}
\Gamma=\left[\begin{array}{cccccccc}
    1 &  0.43 &  0.09  &  0.29 &  0.55  &  0.66 &   0.10 &  -0.01\\
    0.43  &  1 &  -0.54  &  0.11  &  0.37 &  0.25 &  0.51 &  -0.48\\
    0.09 &  -0.54 &  1  & -0.02  &  0.20  &  0.02  & -0.40  &  0.73\\
    0.29 &   0.11&  -0.02 &   1 &  -0.41&  -0.07   &-0.22  &  0.01\\
    0.55  &  0.37  &  0.20  & -0.41  &  1 &   0.43 &   0.31  & 0\\
    0.66 &   0.25  &  0.02 &  -0.07  &  0.43  &  1 &  0.17 &  -0.11\\
    0.10 &   0.51  & -0.40 &  -0.22  &  0.31  &  0.17  &  1  & -0.61\\
   -0.01 &  -0.48 &   0.73  &  0.01  & 0 &  -0.11  & -0.61  &  1\\
\end{array}\right]
\end{equation*}}
In order to compute sensitivity indices, we generate a sample of size  $n=5000$ from this distribution.\\
Here we wish to estimate, for a given operating conditions
and feed vector $c$, the sensitivity
indices of the outputs with respect to the input factors in $\boldsymbol{\theta}$, {\it i.e.}
$$S_i^j=\frac{\textrm{Var}( \E( Y_j | \theta^{i}))}{\textrm{Var}(Y_j)}$$
for $j=1,...,3$ and $i=1,...,8$. Actually, our goal is to identify on
which factor we should make the effort of reducing the uncertainty, by
carrying out new experiments. This factor should be chosen in order to
reduce as much as possible the uncertainty of the outputs.\\
We consider two particular vectors $c_{1}$ and
$c_{2}$ containing the same operating conditions but a different feed
($c_{1}$~: nC4=1 and iC4=0, $c_{2}$~: iC4=1 and nC4=0). We have drawn for each vector
$c_{i}$, $i=1,2$ a sample of
size $n$ from ${\bf Y}$ by Monte-Carlo
simulations, \textit{i.e.} by computing ${\bf Y}_j = \eta(c_i,
\boldsymbol{\theta}_j)$ for $j=1,...,n$. Thus we have a sample from
$({\bf Y},\boldsymbol{\theta})$ for each particular $c_1$ and
$c_{2}$. For instance, the estimates of the sensitivity indices of the third output C3+C5 with the $\hat{T}_{1}$ estimator are given in
Figure 4. Filled bars correspond to $c_{1}$ and empty bars to $c_{2}$.

Note that the estimates given by the $\hat{T}_{2}$ estimator are
similar. These results highlight the behavior of the C3+C5 output
when the feed changes. Indeed when we only use nC4 in the feed
($c_{1}$) the production of C3+C5 is mainly linked to the production
of iC4 by reaction $(1)$. This is confirmed by the importance of parameters $1$ and $6$
in Figure 4 which are the parameters involved in reaction $(1)$. When the
feed only contains iC4 ($c_{2}$), the first
reaction is no longer dominating for
the production of C3+C5, now mainly linked to reaction $(3)$.
Parameters $4$ and $2$ that are the most important in Figure 4 for
$c_{2}$ are connected to reaction $(3)$. We can thus conclude that the
results confirm the expected behavior of the C3+C5 output.\\

\begin{figurehere}
      \includegraphics[width=0.8\linewidth]{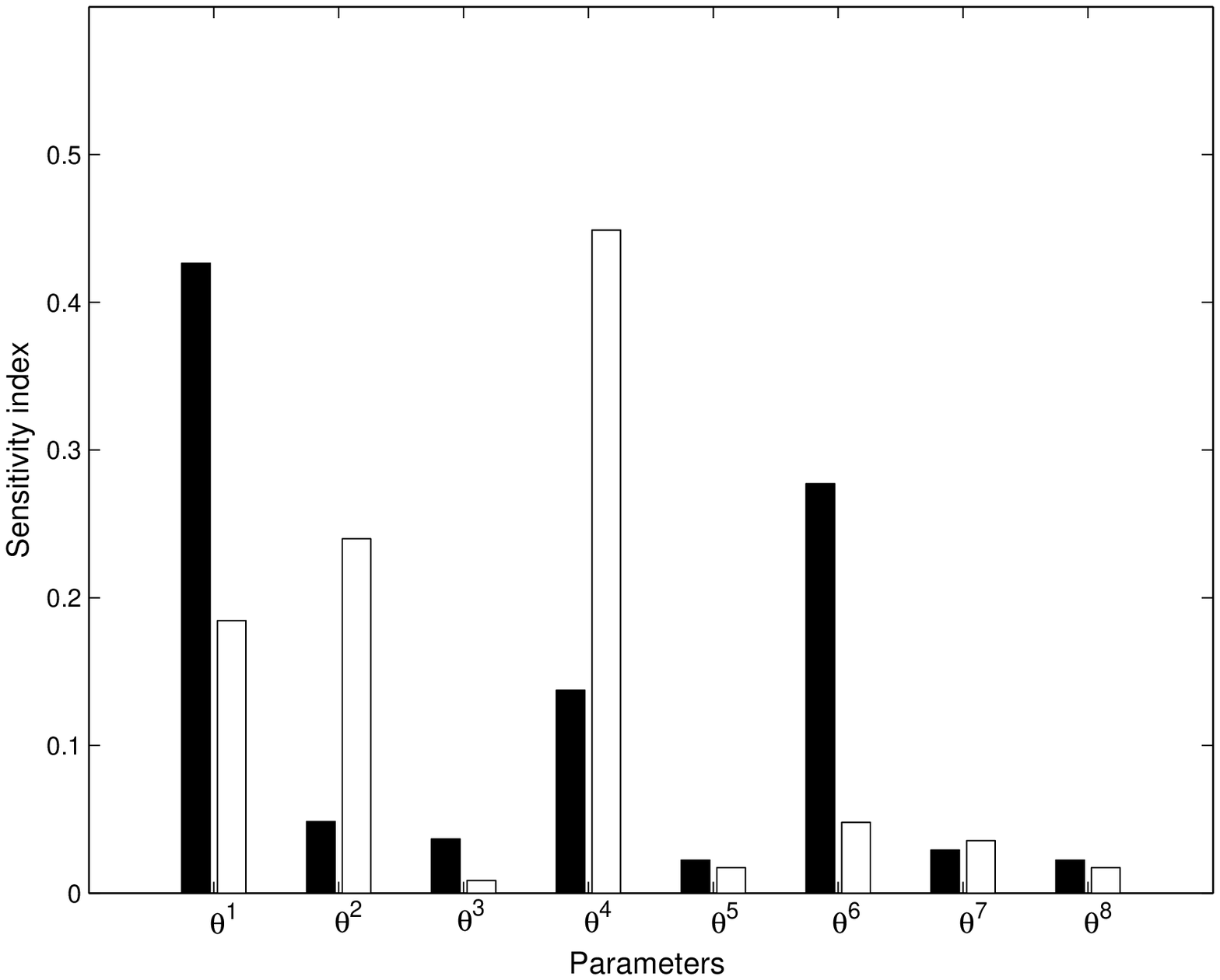}
        \unnumberedcaption{{\small \it Figure 4. Sensitivity indices of the C3+C5 output in the
            isomerization model for the particular conditions $c_{1}$
            (filled bars) and $c_{2}$ (empty bars)}.\\}
\end{figurehere}

We could obviously study the sensitivity indices for the other
outputs, and for other operating conditions. Such a study has been
carried out and showed that the most influent parameters depend on
the operating conditions and the feed. But it also underlined that
each parameter of the model has an influence on at least
one output for at least one operating condition. In this case these sensitivity
indices estimates enlighten the fact that all the parameters are potentially important. A discussion with chemical engineers would then be necessary in order to identify which outputs are most critical for their goals (controlling for instance the first output iC4 which is strongly linked to the octane rate) and would thus help us to choose which input parameters deserve most attention.

\section{DISCUSSION AND CONCLUSION}
The estimation method proposed in this paper is an efficient
way to carry out sensitivity analysis by computing first order
sensitivity indices when the inputs are not independent. The use of
local polynomial estimators is the key point of the estimation procedure.
It guarantees some interesting theoretical properties and ensures good 
qualities to the estimators we have introduced. Beyond these theoretical results,
practical examples also show a good precision for a rather
low computation time. Obviously, higher precision requires higher
calculation time and the user has the possibility to adapt the
estimators, by fixing some hyper-parameter values such as polynomials orders. 

The main advantage of our
estimators is obviously that they only make the assumption that the
marginals are smooth and then require less model runs than classical
sampling methods. Comparing with the Bayesian approach of
\citeasnoun{OO04}, our method has the same philosophy as it uses model
runs to fit a response surface under smoothness assumptions, but we
avoid its numerical integration issue in high dimension.\\
Moreover our approach is appealing for practioners in the
sense that they can see it as a black-box routine, as each
step of the procedure is data-driven once the
user has given the two samples needed for the estimation.\\
Finally, we think that a practitioner willing to carry out a sensitivity analysis should combine different approach to get the most accurate result, for example computing the indices with the method we introduce her and the one of Oakley and O'Hagan. Indeed these two methods are not concurrent but complementary.

Future work will also be based on building multi-outputs
sensitivity indices through multivariate nonparametric regression
techniques.

\section*{ACKNOWLEDGMENTS}
We thank Professor Anestis Antoniadis for very helpful discussions. We
also thank the referees and associate editor for very useful suggestions.

\section*{APPENDIX~: PROOFS OF THEOREMS}

\subsection*{A.1 Assumptions}
We list below all the assumptions we use in the development
of our proofs. Note that the bandwidths $h_{1}$ and $h_{2}$ are by
definition positive real numbers.\\

(A0) As $n\rightarrow \infty$, $h_{1}\rightarrow 0$ and
$nh_{1}\rightarrow \infty$ ;\\

(A1) The kernel $K(.)$ is a bounded symmetric and continuous density function
with finite $7^{th}$ moment ;\\

(A2) $f_{X}(x)>0$ and $\ddot f_{X}(.)$ is bounded in a neighborhood of $x$
where $f_{X}(.)$ denotes the marginal density function of $X$ ;\\

(A3) $\dddot m(.)$ exists and is continuous in a neighborhood of $x$ ;\\

(A4) $\sigma^{2}(.)$ has a bounded third derivative in a neighborhood
of $x$ and $\ddot m(x) \neq 0$ ;\\

(B0) As $n\rightarrow \infty$, $h_{i}\rightarrow 0$ and
$\lim \inf nh_{i}^{4} >0$ for $i=1,2$ ;\\

(B1) The kernel $K(.)$ is a symmetric density function with a bounded
support in $\mathbb{R}$. Further, $|K(x_{1})-K(x_{2})| \leq c
|x_{1}-x_{2}|$ for $x_{1}, x_{2} \in \mathbb{R}$ ;\\

(B2) The marginal density function $f_{X}(.)$ satisfies $f_{X}(x)>0$ and $|f_{X}(x_{1})-f_{X}(x_{2})| \leq c
|x_{1}-x_{2}|$ for $x_{1}, x_{2} \in \mathbb{R}$ ;\\  

(B3) $\E(Y^{4}) < \infty$ ;\\

(B4) $\sigma^{2}(x)>0$ and the function $\E(Y^{k}|X=.)$
is continuous at $x$ for $k=3,4$. Further, $\dddot m(.)$ and $\dddot
\sigma^{2}(.)$ are uniformly continuous on an open set containing the
point $x$ ;\\

(C0) $f_{X}(.)$ has compact support $[a,b]$\\

Assumptions (A0) and (B0) are standard ones in kernel
estimation theory. Some classical considerations on MSE or MISE (Mean
Integrated Squared Error) lead to theoretical optimal constant
bandwidths of order $n^{-1/5}$.

Assumptions (A1) and (B1) are directly satisfied by commonly
used kernel functions. We can note that they require a kernel with
bounded support, but this is only a technical assumption for brevity
of proofs. For example, the Gaussian kernel can be used.

The assumption $f_{X}(x)>0$ in (A2) and (B2) simply ensures that the
experimental design is rich enough. The fact that (A2) also requires
$\ddot f_{X}(.)$ to be bounded in a neighborhood of $x$ is
natural. The Lipschitz condition on $f$ in (B2) is directly satisfied
if $f$ is sufficiently regular and with compact support.

Assumptions (A3), (A4), (B3) and (B4) are natural and ensure
sufficient regularity to the conditional moments.

Assumption (C0) is made to make the presentation easier. It
can be relaxed by means of the conventional truncation techniques
used in real cases (\citeasnoun{MS82}). Nevertheless in
practice, the input factors considered in sensitivity analysis are
bounded and so have densities with compact support.

\subsection*{A.2 Proof of Theorem 1}
This theorem is a direct consequence of the asymptotic
behavior of the bias and variance in local linear regression. 

Under assumptions (A0)-(A4), \citeasnoun{FGH96} established
that for a given kernel $K(.)$ 
\begin{equation}
\E_{\mathbb{X}}(\hat{m}(x))=m(x)+\frac{1}{2} \mu_{2} \ddot
m(x)h_{1}^{2}+o_{P}(h_{1}^{2})    \label{FAN1}
\end{equation}
and
\begin{equation}
\textrm{Var}_{\mathbb{X}}(\hat{m}(x))=\frac{\nu_{0}\sigma^{2}(x)}{f_{X}(x)n
  h_{1}}+o_{P}(h_{1}^{2})         \label{FAN2}
\end{equation}
where $\displaystyle{\mu_{k}=\int u^{k} K(u)du}$ and
$\displaystyle{\nu_{k}=\int u^{k} K^{2}(u)du}$. Now as the estimator $\hat{T}_1$ is
$$ \hat{T}_{1}=\frac{1}{n'-1}
\sum_{j=1}^{n'} \left( \hat{m}(\tilde{X}_{j})-\hat{\bar{m}}
\right)^{2}$$
we can write
$$\hat{T}_{1}=\frac{1}{n'-1}\sum_{j=1}^{n'}(Z_{j}-\bar{Z})^{2}$$
where $(Z_{j})_{j=1,...,n'}:=
(\hat{m}(\tilde{X}_{j}))_{j=1,\ldots,n'}$ and
$\displaystyle{\bar{Z}=\frac{1}{n'}\sum_{j=1}^{n'}Z_{j}}$. By
conditioning on the predictors $\mathbb{X}$, the sample $(Z_{j}|\mathbb{X})_{j=1,...,n'}$ is an
i.i.d. sample distributed as $Z_{1}|\mathbb{X}$ and the conditional
bias of $\hat{T}_{1}$ can then be obtained through the classical formula for the
empirical estimator of the variance :
$$\E_{\mathbb{X}}(\hat{T}_{1})=\textrm{Var}_{\mathbb{X}}(Z_{1})=\E_{\mathbb{X}}(Z_{1}^{2})- \E_{\mathbb{X}}(Z_{1})^{2}.$$
Note that we can also compute its variance
\begin{equation*}
\textrm{Var}_{\mathbb{X}}(\hat{T}_{1})=\frac{1}{n'} \left( \E_{\mathbb{X}}((Z_{1}-\E_{\mathbb{X}}(Z_{1}))^{4})
  -\frac{n'-3}{n'-1} \left( \textrm{Var}_{\mathbb{X}}(Z_{1})\right)^{2}
\right) \label{variance}
\end{equation*}
even though we do not use this result here (see Remark 1.).

\noindent As $\tilde{X}$ is independent of $X$ and $Y$, we write
\begin{eqnarray*}
  \E_{\mathbb{X}}(Z_{1}^{2})&=&\int \E_{\mathbb{X}}(\hat{m}(x)^{2})f_{\tilde{X}}(x)dx \\
&=& \int \left( \textrm{Var}_{\mathbb{X}}(\hat{m}(x)) +
  \E_{\mathbb{X}}(\hat{m}(x))^{2}\right) f_{X}(x)dx.
\end{eqnarray*}
Considering assumptions (A3), (A4) and (C0) we then get using
(\ref{FAN1}) and (\ref{FAN2}), in a similar way as for the standard
MISE evaluation, 
\begin{eqnarray*}
  \E_{\mathbb{X}}(Z_{1}^{2})&=& \int m(x)^{2}f_{X}(x)dx + \frac{\nu_{0}}{nh_{1}}
  \int \sigma^{2}(x)dx \\
&& + \mu_{2}h_{1}^{2} \int m(x)\ddot m(x)f_{X}(x)dx + o_{P}(h_{1}^{2})\\
\end{eqnarray*}
\noindent and by the same arguments we also have
$$\E_{\mathbb{X}}(Z_{1})= \int m(x)f_{X}(x)dx +\frac{1}{2}\mu_{2}h_{1}^{2}\int
\ddot m(x)f_{X}(x)dx + o_{P}(h_{1}^{2}),$$
which finally leads to
\begin{eqnarray*}
  \E_{\mathbb{X}}(\hat{T}_{1})&=&\E_{\mathbb{X}}(Z_{1}^{2})- \E_{\mathbb{X}}(Z_{1})^{2}\\
&=& \textrm{Var}(\E(Y|X)) \\
&& + \mu_{2}h_{1}^{2} \left[ \int m(x)\ddot m(x) f_{X}(x) dx \right.\\
&& \left. - \left(
    \int m(x)f_{X}(x)dx\right)\left( \int \ddot m(x)f_{X}(x)dx\right)
\right]\\
&& + \frac{\nu_{0}}{nh_{1}} \int \sigma^{2}(x)dx + o_{P}(h_{1}^{2})\\
&=& \textrm{Var}(\E(Y|X))+M_{1}h_{1}^{2}+\frac{M_{2}}{nh_{1}}+o_{P}(h_{1}^{2})
\end{eqnarray*}\\
where 
\begin{eqnarray*}
M_{1} &=& \mu_{2} \left[ \int m(x)\ddot m(x) f_{X}(x) dx \right.\\
&& \left.- \left(
    \int m(x)f_{X}(x)dx\right)\left( \int \ddot m(x)f_{X}(x)dx\right)
\right]
\end{eqnarray*}
and
$$M_{2}=\nu_{0} \int \sigma^{2}(x)dx.$$

\subsection*{A.3 Proof of Theorem 2}
Similarly we first recall asymptotic results for the
residual-based estimator of the conditional variance. 

Under assumptions (B0)-(B4) \citeasnoun{FY98} showed that
\begin{equation*}
\E_{\mathbb{X}}(\hat{\sigma}^{2}(x))=\sigma^{2}(x)+\frac{1}{2} \mu_{2} \ddot
\sigma^{2}(x)h_{2}^{2}+o_{P}(h_{1}^{2}+h_{2}^{2})    \label{FAN3}
\end{equation*}
and
\begin{equation*}
\textrm{Var}_{\mathbb{X}}(\hat{\sigma}^{2}(x))=\frac{\nu_{0}\sigma^{4}(x)\lambda^{2}(x)}{f_{X}(x)n
  h_{2}}+o_{P}\left(\frac{1}{\sqrt{nh_{2}}}\right)    \label{FAN4}
\end{equation*}
where $\lambda^{2}(x)=\E((\epsilon^{2}-1)^{2}|X=x)$ and $\mu_{2}$ and
$\nu_{0}$ are as defined above. The estimator $\hat{T}_{2}$ can be written as
$$\hat{T}_{2}=\frac{1}{n'}\sum_{j=1}^{n'} U_{j}$$
where
$(U_{j})_{j=1,...,n'}:=(\hat{\sigma}^{2}(\tilde{X}_{j}))_{j=1,\ldots,n'}$.
As in the proof of Theorem 1, we then get the conditional bias and
variance of $\hat{T}_{2}$~:
$$\E_{\mathbb{X}}(\hat{T}_{2})=\E_{\mathbb{X}}(U_{1})$$
and
$$\textrm{Var}_{\mathbb{X}}(\hat{T}_{2})=\frac{1}{n'}\textrm{Var}_{\mathbb{X}}(U_{1}).$$
As $\tilde{X}$ is independent of $X$ and $Y$, we have
$$\E_{\mathbb{X}}(U_{1})=\int \E_{\mathbb{X}}(\hat{\sigma}^{2}(x)) f_{\tilde{X}}(x)dx.$$
Considering assumptions (B4) and (C0) as in the proof of
Theorem 1 we then get
\begin{eqnarray*}
\E_{\mathbb{X}}(\hat{T}_{2}) &=& \E(\textrm{Var}(Y|X))+\frac{1}{2}\mu_{2}h_{2}^{2}\int
\ddot \sigma^{2}(x)f_{X}(x)dx\\
&& + o_{P}(h_{1}^{2}+h_{2}^{2})\\
&=& \E(\textrm{Var}(Y|X))+ V_{1}h_{2}^{2}+ o_{P}(h_{1}^{2}+h_{2}^{2})
\end{eqnarray*}
where
$$V_{1}=\frac{1}{2}\mu_{2}\int
\ddot \sigma^{2}(x)f_{X}(x)dx $$
and using the same arguments 
\begin{eqnarray*}
  \textrm{Var}_{\mathbb{X}}(\hat{T}_{2}) & = & \frac{1}{n'} \left\{
    \E(\textrm{Var}(Y|X)^{2})\right.\\
& &  + \mu_{2} h_{2}^{2} \int \sigma^{2}(x) \ddot
\sigma^{2}(x)f_{X}(x)dx  \\
&  & - \mu_{2}h_{1}^{2} \left( \int \ddot \sigma^{2}(x)f_{X}(x)dx
\right) \left( \int \sigma^{2}(x)f_{X}(x)dx\right) \\
&  & \left. + \frac{\nu_{0}}{nh_{2}} \int \sigma^{4}(x)
  \lambda^{2}(x)dx \right. \\
&& \left. + o_{P}\left(h_{1}^{2}+h_{2}^{2}+ \frac{1}{\sqrt{nh_{2}}}\right)
\right\} \\
&=& \frac{1}{n'} \left\{ \E(\textrm{Var}(Y|X)^{2}) + V_{2} h_{2}^{2}+
  V_{3}h_{1}^{2} +\frac{V_{4}}{nh_{2}} \right.\\
&& \left. + o_{P}\left(h_{1}^{2}+h_{2}^{2}+ \frac{1}{\sqrt{nh_{2}}} \right) \right\}
\end{eqnarray*}
where
$$V_{2}=\mu_{2} \int \sigma^{2}(x) \ddot
\sigma^{2}(x)f_{X}(x)dx,$$
$$V_{3}=-\mu_{2} \left( \int \ddot \sigma^{2}(x)f_{X}(x)dx
\right) \left( \int \sigma^{2}(x)f_{X}(x)dx\right),$$
$$V_{4}=\nu_{0}\int \sigma^{4}(x) \lambda^{2}(x)dx.$$

\bibliographystyle{agsm}
\bibliography{article}

\end{multicols}
  
\end{document}